\documentclass{elsart}

\usepackage{amssymb}
\usepackage{graphicx}

\journal{Physica A}

\begin{document}

\begin{frontmatter}


\title{Microscopic explanation of non-Debye relaxation for heat transfer}

\author{Agata Fronczak \corauthref{cor}},
\corauth[cor]{Corresponding author. {\it E-mail address}:
agatka@if.pw.edu.pl}\author{Piotr Fronczak, Janusz A. Ho\l yst}

\address{Faculty of Physics and Center of Excellence for
Complex Systems Research, Warsaw University of Technology,
Koszykowa 75, PL-00-662 Warsaw, Poland}

\begin{abstract}
We give a microscopic explanation of both Debye and non-Debye
thermalization processes that have been recently reported by Gall
and Kutner \cite{Kutner2005}. Due to reduction of the problem to
first passage phenomena we argue that relaxation functions $f(t)$
introduced by the authors directly correspond to survival
probabilities $S(t)$ of particles in the considered systems. We
show that in the case of broken ergodicity (i.e. in the case of
mirror collisions) the survival probability decays as a power law
$S(t)=\tau/t$.
\end{abstract}

\end{frontmatter}

\section{Introduction: motivation and aims} \label{intro}

A few decades ago, it was widely accepted that the relaxation of a
typical macroscopic observable $\mathcal{A}$ towards its
equilibrium value is described by exponential function
\begin{equation}\label{ftexp}
f(t)=\frac{\Delta\mathcal{A}(t)}{\Delta\mathcal{A}(0)}=e^{-t/\tau_0},
\end{equation}
where $t$ represents time and $\Delta\mathcal{A}(t)=
\mathcal{A}(t)-\mathcal{A}(\infty)$. On that score, probably the
best-known example is the Newton's law of cooling where the
relaxation function $f(t)$ refers to thermalization process (i.e.
$\mathcal{A}=T$). Recently, however, it was discovered that a
number of decay phenomena observed in nature obeys slower
non-exponential relaxation (c.f
\cite{book_nonDebey,PRLchaos,Koks1,Koks2,PRLquant})
\begin{equation}\label{ftsf}
f(t)=\left(\frac{t}{\tau}\right)^{-d}.
\end{equation}
It was also noticed that non-Debye relaxation is usually observed
in systems which violate the ergodicity condition. In such
systems, the lack of ergodicity results from long-range
interactions, microscopic memory effects or from (multi)fractal
structure of phase space. It was argued that such systems are well
described by the so-called nonextensive statistics introduced by
Tsallis \cite{JSPTsallis,book_Tsallis}. The {\it conjectured}
relaxation function for such systems \cite{PhysDTsallis,Weron2004}
is given by the $q$-exponential decay \footnote{As a matter of
fact, the only satisfactory proof of the property (\ref{ftqexp})
exists for such systems in which nonextensivity arises from
intrinsic fluctuations of some parameters describing the system's
dynamics \cite{PRLWilk,PhysLettAWilk}.}
\begin{equation}\label{ftqexp}
f(t)=e_q^{-t/\tau_q}=\left[1+(q-1)\frac{t}{\tau_q}\right]^{1/(1-q)}.
\end{equation}
which is equivalent to the formula (\ref{ftexp}) for $q\simeq 1$,
whereas for $q>1$ it coincides with (\ref{ftsf}).

In the context of the ongoing discussion on possible relations
between nonergodicity and nonextensivity, the phenomenon of
non-Debye relaxation that has been recently reported by Gall and
Kutner \cite{Kutner2005} seems to be particularly interesting (see
also \cite{KutnerCarnot}). The authors have numerically studied a
simple molecular model as a basis of irreversible heat transfer
trough a diathermic partition. The partition has separated two
parts of box containing ideal point particles (i.e. ideal gases)
that have communicated only through this partition (see
Fig.~\ref{fig1}a). The energy transfer between the left and
right-hand side gas samples has consisted in equipartition of
kinetic energy of all outgoing particles colliding with the
partition during a given time period. The authors have analysed
and compared two essentially different cases of the system's
dynamics:
\begin{enumerate}
\item[i.] the first case, where the border walls of the box and
the diathermic partition have randomized the direction of the
motion of rebounding particles, and
\item[ii.] the case, where mirror collisions of particles with
the border walls and the partition have been
considered.
\end{enumerate}
They have found that although the mechanism of heat transfer has
been analogous in both cases the long-time behaviour of both
thermalization processes has been completely different. In the
first case (i.) ordinary Debye relaxation of the system towards
its equilibrium state has been observed
\begin{equation}\label{Kutexp}
\Delta T(t)\sim e^{-t/\tau_0},
\end{equation}
where $\Delta T(t)=T_1(t)-T_2(t)$ is the temperature difference
between both gas samples, while in the second case (ii.) the
power-law decay has been noticed
\begin{equation}\label{Kutsf}
\Delta T(t)\sim\frac{\tau}{t}.
\end{equation}

In order to describe the phenomenon of the non-Debye relaxation
Gall and Kutner \cite{Kutner2005} have derived an extended version
of the thermodynamic Fourier-Onsager theory \cite{FOT1,FOT2} where
heat conductivity was assumed to be time-dependent quantity. The
authors have argued that from the microscopic point of view the
non-Debye relaxation results from the fact that the gas particles
always move along fixed orbits in the case (ii.). They have also
argued that this regular motion may be considered as nonergodic,
violating the molecular chaos hypothesis (Boltzmann, 1872).

In this paper we propose a more rigorous microscopic explanation
of both Debye and non-Debye thermalization processes reported by
Gall and Kutner.

\section{Microscopic model for non-Debye heat transfer}\label{Micro}

\begin{figure}
\begin{center}
\includegraphics*[width=6cm]{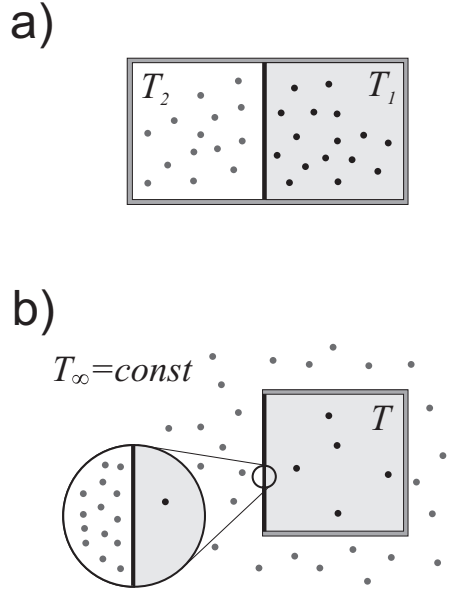}
\end{center}
\caption{(a) {\it Original experimental system} \cite{Kutner2005}.
Two gas samples exchanging heat through a diathermic partition.
(b) {\it Reduced model system.} Gas sample in thermal contact with
a huge heat reservoir at constant temperature $T_\infty$ (detailed
description is given in the text.)} \label{fig1}
\end{figure}

In the paper \cite{Kutner2005}, the authors have analysed
two-dimensional systems consisting of two gas samples of
comparable size (see Fig.~\ref{fig1}a). Here, due to analytical
simplicity we assume that one gas sample is significantly larger
and denser than the second one i.e. the larger sample may be
referred to as a heat reservoir with constant temperature
$T_{\infty}=const$ (see Fig.~\ref{fig1}b). We also assume that the
smaller sample is confined in a square box of linear size $l$ and
its initial temperature equals $T_0$. It is natural to expect that
thanks to the existence of the diathermic partition the
temperature of the smaller sample will tend to the reservoir
temperature $T(t)\rightarrow T_{\infty}$ in the course of time.

At the moment, before we analytically justify the relaxation
functions (\ref{Kutexp}) and (\ref{Kutsf}) let us recall crucial
assumptions of the numerical experiment performed by Gall and
Kutner \cite{Kutner2005}. First, the authors have defined the
temperature of the given gas sample $T(t)$ as proportional to the
average kinetic energy of all particles in the sample
\begin{equation}\label{Tdef}
kT(t)=\frac{1}{N}\sum_{i=1}^{N}\varepsilon_i(t).
\end{equation}
Second, they have assumed a monoenergetic energy distribution
function $P(\varepsilon) =\delta(\varepsilon-kT_i)$ as the initial
condition for each gas sample $i=1,2$ (see Fig.~\ref{fig1}a). In
fact, since the applied thermalization mechanism evens out kinetic
energies of all particles colliding with the diathermic partition
at a given time not only initial but also final (i.e. equilibrium)
energy distributions are monoenergetic.

Now, having in mind the above assumptions, one can simply conclude
that during the heat transfer occurring in the system presented at
Fig.~\ref{fig1}b the particles of the smaller and thinner gas
sample get the final energy immediately after {\it the first
collision} with the diathermic partition \footnote{Gall and Kutner
have proved that the systems presented at Fig. \ref{fig1}a possess
a similar feature for asymptotic times (cf. Eq. (45) in
\cite{Kutner2005}).} i.e. $\varepsilon_0\rightarrow
\varepsilon_{\infty}$, where $\varepsilon_0=kT_0$ and
$\varepsilon_\infty=kT_\infty$ (\ref{Tdef}). It is possible due to
the existence of the huge and dense heat reservoir which causes
that the number of particles with energy $\varepsilon_\infty$
colliding with the diathermic partition at a given time is
overwhelmingly larger than the number of particles with the energy
$\varepsilon_0$ which at the same time collide with the partition
from the other site (see the inset in Fig.~\ref{fig1}b).

The above considerations allow us to write the temperature
difference between both gas samples in the following way
\begin{equation}\label{af1}
\Delta T(t)=T(t)-T_{\infty}=\frac{N(t)}{N}(T_0-T_\infty),
\end{equation}
where $N(t)$ is the number of particles of the smaller sample
which have not hit the diathermic partition by time $t$. Now, one
can see that the relaxation function $f(t)$ of the considered
systems is equivalent to the survival probability $S(t)=N(t)/N$
\begin{equation}\label{af2}
f(t)=\frac{T(t)-T_{\infty}}{T_0-T_{\infty}}\equiv S(t).
\end{equation}

The last formula makes us possible to reduce the phenomena of
Debye and non-Debye relaxations to the first passage processes
\cite{book_Redner,Risken,Gardiner}. In this sense, the case (i.)
of rough border walls directly corresponds to the problem of
diffusing particles in a finite domain with an absorbing boundary.
The survival probability $S(t)$ typically decays exponentially
with time for such systems \cite{book_Redner}. That is the reason
way the thermalization process characterizing the case (i.) is
equivalent to Debye relaxation (see Eq.~(\ref{Kutexp})). In the
next paragraph we show that the case (ii.), where mirror bouncing
walls and absorbing diathermic partition are taken into account,
is indeed characterized by the power law decay of the survival
probability $S(t)$ (see Eq.~(\ref{Kutsf})).

\begin{figure}
\begin{center}
\includegraphics*[width=12cm]{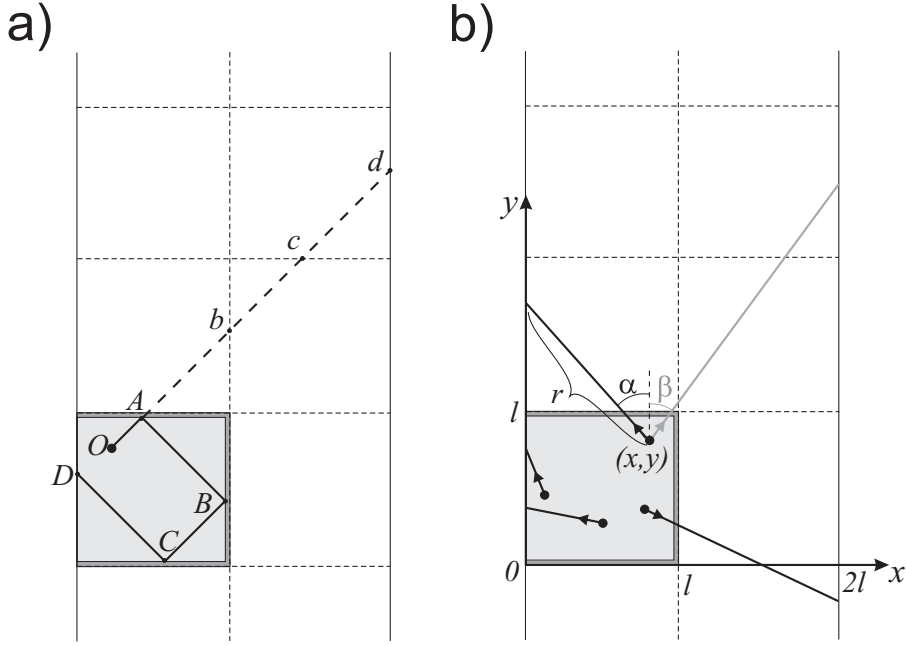}
\end{center}
\caption{Ideal point particles in the box with mirror border walls
(detailed description is given in the text).} \label{fig2}
\end{figure}

In order to achieve the claimed scale-free decay of the survival
probability $S(t)$ one has to find the so-called first passage
probability $F(t)$ (i.e. the probability that a particle of the
considered gas sample hits the diathermic partition for the first
time at the time $t$)
\begin{equation}\label{af3}
S(t)=1-\int^{t}_0F(t')dt'.
\end{equation}
At the moment, let us remind that before the collision with the
diathermic partition each particle has the same velocity $v_{0}$
(i.e. $mv_0^2/2=\varepsilon_0=kT_0$), thus the distribution $F(t)$
can be simply calculated from the particle path length
distribution $\widetilde{F}(r)$ i.e.
\begin{equation}\label{af4}
F(t)=\widetilde{F}(r)\left|\frac{dr}{dt}\right|=\widetilde{F}(v_0t)v_0,
\end{equation}
where $r=v_0t$. Now, due to the symmetry of the considered problem
(that is due to the equivalence of paths $0\rightarrow A
\rightarrow B\rightarrow C\rightarrow D$ and $0\rightarrow A
\rightarrow b\rightarrow c\rightarrow d$, see Fig.~\ref{fig2}a)
one can deduce the following relation
\begin{equation}\label{af5}
\widetilde{F}(r)=\frac{1}{2}P(\alpha)\left|\frac{d\alpha}{dr}\right|+
\frac{1}{2}P(\beta)\left|\frac{d\beta}{dr}\right|,
\end{equation}
where $0\leq \alpha,\beta\leq\pi/2$ and respectively
$\sin\alpha=x/r$ whereas $\sin\beta=(2l-x)/r$ (see
Fig.~\ref{fig2}b). The last formula expresses the fact that
particles can either move to the left (i.e. towards the diathermic
partition) or to the right (i.e. towards mirror reflection of the
partition). Next, assuming uniform initial conditions
$P(\alpha)=P(\beta)=2/\pi$ in the long time limit (i.e. for small
angles when $\sin\alpha\simeq\alpha$ and $\sin\beta\simeq\beta$)
one obtains
\begin{equation}\label{af6}
F(t)=\frac{2l}{\pi v_0}\;t^{-2}.
\end{equation}
Finally, using the relation (\ref{af3}) one gets the desired
power-law decay of the survival probability which justifies the
non-Debye thermalization process (\ref{Kutsf})
\begin{equation}\label{af7}
S(t)=\frac{\tau}{t},\;\;\;\;\;\mbox{where}
\;\;\;\;\;\tau=\frac{2l}{\pi v_0}.
\end{equation}

\begin{figure}
\begin{center}
\includegraphics*[width=10cm]{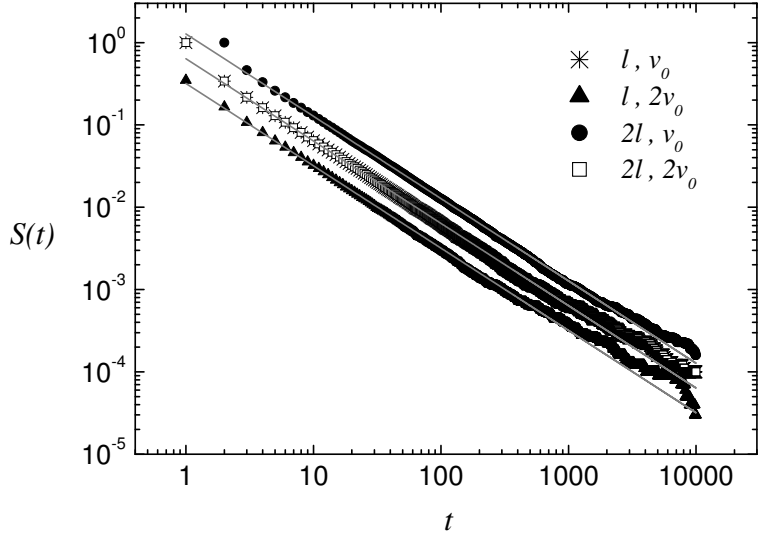}
\end{center}
\caption{Survival probability $S(t)$ against time $t$ in systems
presented at Fig.~\ref{fig2}. Points correspond to results of
numerical simulations whereas solid lines represent theoretical
prediction of the formula (\ref{af7}).} \label{fig3}
\end{figure}

We have numerically verified the last relation for a few different
values of both the initial velocity $v_0$ and the box size $l$. In
all the considered cases we have obtained very good agreement of
recorded survival probabilities with the formula (\ref{af7}) (see
Fig.~\ref{fig3}).

\section{Summary and concluding remarks}

In this paper we have given a microscopic explanation of Debye and
non-Debye thermalization processes that have been recently
reported by Gall and Kutner \cite{Kutner2005}. The authors have
studied a simple molecular mechanism of heat transfer between two
comparable gas samples. Owing to analytical simplicity we have
reduced the problem to one gas sample being in thermal contact
with the huge and dense heat reservoir at constant temperature.
For the case we have shown that the thermalization mechanism
described by Gall and Kutner can be reduced to first passage
phenomena. Taking advantage of the idea we have found an
analytical justification for both exponential (\ref{Kutexp}) and
non-exponential (\ref{Kutsf}) relaxation functions observed in
\cite{Kutner2005}.

\section{Acknowledgments}

We thank Prof. Ryszard Kutner for stimulating discussions. This
work was financially supported by internal funds of the Faculty of
Physics at Warsaw University of Technology (Grant No.
503G10500021005). A.F. also acknowledges the financial support of
the Foundation for Polish Science (FNP 2005).

\end{document}